# Evaluation of a gate capacitance in the sub-aF range for a chemical field-effect transistor with a silicon nanowire channel

Nicolas Clément, Katsuhiko Nishiguchi, Akira Fujiwara, & Dominique Vuillaume

*Abstract*— An evaluation of the gate capacitance of a field-effect transitor (FET) whose channel length and width are several ten nanometer, is a key point for sensors applications. However, experimental and precise evaluation of capacitance in the aF range or less has been extremely difficult. Here, we report an extraction of the capacitance down to 0.55 aF for a silicon FET with a nanoscale wire channel whose width and length are 15 and 50 nm, respectively. The extraction can be achieved by using a combination of four kinds of measurements: current characteristics modulated by double gates, random-telegraph-signal noise induced by trapping and detrapping of a single electron, dielectric polarization noise, and current characteristics showing Coulomb blockade at low temperature. The extraction of such a small gate capacitance enables us to evaluate electron mobility in a nanoscale wire using a classical model of current characteristics of a FET.

## I. INTRODUCTION

SEMICONDUCTOR nanowire (NW) field-effect transistors (FETs) are of great interest for future electronic devices with high performance[1-3], and low power consumption[4], as well as chemical sensors[5-7]. Meanwhile, an evaluation of a gate capacitance of such a NW FET is extremely important for an extraction of intrinsic carrier mobility in the NW from the viewpoint of device applications and for an estimation of charge detected by the FET used as the chemical sensor. Interestingly, carrier mobility in a p-type Si NW was evaluated[1] to be as high as 1350 $cm^2/Vs$, an order of magnitude larger than that in a bulk Si, from a capacitance estimated geometrically. However, since following works have reported the mobility in longer Si NWs close to bulk values[8,9], more precise evaluation of the capacitance to estimate an accurate mobility is required. It is extremely challenging due to its small value in the aF range or less, which is usually masked by much larger parasitic capacitances in the pF range or larger. Although an optimization of a capacitance bridge allowed direct evaluation of capacitance (300 aF) of a 700-nm-length nanowire, its measurement was carried out at 200 K[8]. Moreover, an evaluation of a capacitance of several-ten-nanometer-length NW has not been reached. Room-temperature measurement was carried out using hundreds of 300-nm-length Si NWs connected in parallel in order to increase the capacitance, and to evaluate the averaged capacitance among a lot of Si NWs[9]. Recently, new emerging techniques for an evaluation of capacitances in the aF range were reported: a scanning microwave microscope[10] and noise analysis[11,12], However, each technique has inevitably uncertainties in the aF range, which urges extra analysis of those feasibility by comparison with several independent techniques.

Recently, ultra-low-noise Si NW chemical field-effect transistors (CHEMFETs) with elementary charge sensitivity have been demonstrated[13]. It opens the door to a metrological approach for the electrical detection of single ions and molecules. Here, we use these ultra-low-noise Si NW CHEMFETs to evaluate gate capacitances using four kinds of measurements: current (I) characteristics modulated by double gates, random-telegraph signal (RTS) induced by trapping and detrapping of a single electron, dielectric polarization (DP) noise, and current characteristics showing a behavior of a single-electron transistor at low temperature.

## II. EXPERIMENTAL

### A. Device description

The CHEMFETs were analyzed in two ways to modulate current characteristics: a single gate (back gate) and double gates (electrolytic and back gates) as shown in Figs. 1a and 1b, respectively. The devices were fabricated using an undoped silicon-on-insulator (SOI) wafer whose oxide thickness tox between the channel and substrate used for a back gate was 400 nm. The channel formed on the SOI layer was locally constricted for the Si NW and oxidized for the formation of an upper oxide layer whose thickness was 40 nm. Thickness tsi, width W, and length L of the Si NW were approximately 15, 15, and 50 nm, respectively. The measured current characteristics were dominated by the constriction, i.e., Si NW[14]. Such a small wire channel makes the MOSFET useful as a high-charge-sensitivity electrometer with single-electron resolution at room temperature[13-16]. Next, platinum electrodes and 300 µm thick micro-baths made of polymerized resist were formed for an electrolytic gate [13]. An optical and atomic force microscope image of the device are shown in Figs. 1d and 1e, respectively. Single- and double-gate operation of the CHEMFET can be carried out when the micro-bath is filled with $N_2$ and aqueous electrolyte, respectively. Previously we have shown that the capacitance (Cug) between the electrolytic gate and channel is determined only by the oxide surrounding the Si NW. Indeed, Cug is composed of an electrolytic double layer capacitance in series the upper oxide capacitance. The latter one dominates Cug due the higher dielectric constant of water compared to $SiO_2$ and to the thick upper oxide layer (40 nm) [13]. Therefore, this study can be adapted to NW FET with any type of double gates, e.g. gates made of polycrystalline Si and metal.

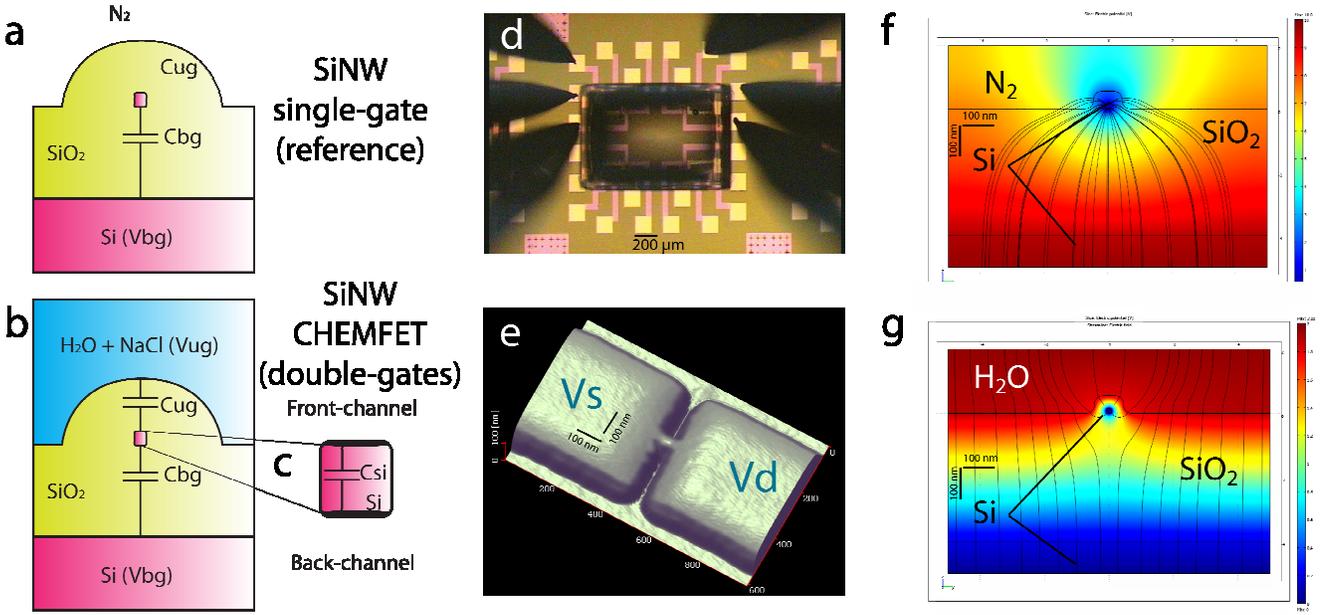

Fig. 1. Device structure and finite element analysis (FEA) simulation (a) Cross-section views of our structure in the single-gate configuration (b) cross-section view of the our structure in double-gate configuration where the electrolyte acts as an upper-gate. The capacitive network including Cbg, Cug and Csi is represented. (c) Zoom on the SiNW which indicates the front and back-channel for electrons. (d) Top view of the device: optical image of the liquid-gated structure (e) AFM image of the SiNW with contact pads. (f) Cross section views of electric field lines and isopotentials simulated by FEA (no upper gate with Vbg = 10 V, Vd=Vs = 0 V) and (g) (liquid-gated with Vug = 2V, Vbg = Vd=Vs = 0 V).

*B. Finite Element Analysis simulaion (Comsol©)*

For basic information, we first evaluate the gate capacitances by a finite element analysis (FEA) simulation using COMSOL©. Figs. 1f and 1g show the cross-sectional views of electrical potential in the single- or double-gate configuration, respectively. Cbg in the single-gate configuration and Cug in the double-gate configuration are evaluated to be 0.76 aF and 3.1 aF, respectively. This method is useful for rough estimation of the capacitance.

*C. Double-gates configuration*

The first experimental approach takes advantage of a benefit originating from the double-gate configuration. In the double-gate operation of the CHEMFET, electron channels are induced either at a top or bottom of the Si NW, hereafter referred to front or back channels, according to upper-gate voltage (Vug) applied to the electrolytic gate and back-gates voltage (Vbg) applied to the back gate (Fig. 1c), respectively, as shown in the insets of Fig. 2b. In this case, we must consider capacitance Csi formed between the top and bottom of the Si NW as shown in Fig. 1b. Although Csi can be basically given by $\varepsilon_{Si}WL/tsi$, where $\varepsilon_{Si}$ is the dielectric constant of silicon, we must arrange it according to the channel configuration [17]. In the back channel configuration, we must consider that there are Csi and Cug between the upper gate and electron channel, which means that effective capacitance Cugsi between them can be given by Cug.Csi/(Cug+Csi). Although the front-channel configuration requires the same consideration, we can ignore it because Cbg is much smaller than Csi. Taking $\Gamma_f$ given by -Cbg/Cug in the front-channel configuration and $\Gamma_b$ given by -Cbg/Cugsi in the back-channel configuration, we can get the following equations that make no assumption for Cbg and Cug.

$$Cugsi \approx \frac{\varepsilon_{Si}WL}{tsi}\left[1-\frac{\Gamma_f}{\Gamma_b}\right], \; Cug \approx \frac{\varepsilon_{Si}WL}{tsi}\left[\frac{\Gamma_b}{\Gamma_f}-1\right], \; Cbg \approx \frac{\varepsilon_{Si}WL}{tsi}\left[\Gamma_b-\Gamma_f\right] \quad (1)$$

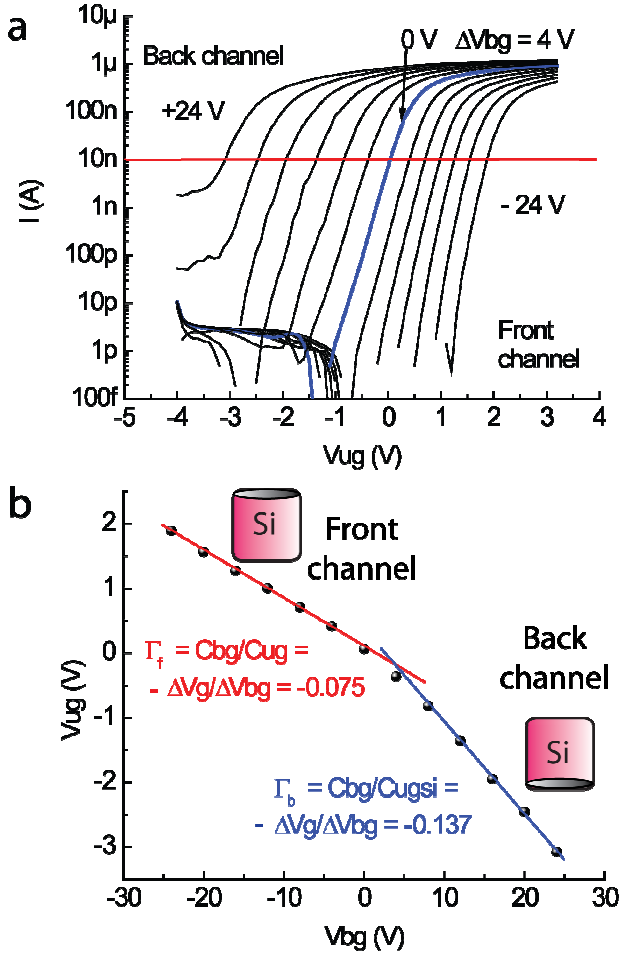

Fig. 2. Estimation of capacitances from double-gates electrical measurements(a) I-Vg for Vbg ranging from -24 V to + 24 V by +4 V steps. (b) (Vg,Vbg) at 10 nA is extracted from (a).

$\Gamma_f$ and $\Gamma_b$ can be obtained experimentally from I-Vug characteristics measured at different Vbg values as shown in Fig. 2a. Fig. 2b shows change in Vug when I becomes 10 nA at various Vbg's. The two different slopes were confirmed; one slope of - 0.075 at positive Vug corresponds to $\Gamma_f$, and the other of -0.137 at negative Vug corresponds to $\Gamma_b$. Then, we can evaluate Csi, Cug, and Cbg to be 5.27, 4.36, and 0.32 aF, respectively, using Eq. 1. The double-gate operation provides the precise evaluation of a ratio between Cug and Cbg. The estimation of the Si NW size is also accurate. Therefore we can derive capacitances using Eq. 1.

*C. Random Telegraph Signal*

The second experimental approach is based on a discrete change ΔI in current called RTS induced by a single electron trapped and detrapped between the channel and oxide layer. Typical examples of the RTS are shown in Fig. 3a and 3b. It should be emphasized that the shift of I-Vug characteristics induced by electron trapping does not depend on a trap position in the oxide[11]. This is because a distance between the channel and trap is much smaller than oxide thickness and also because a Debye screening length in an undoped Si NW is much larger than device dimensions. Therefore, ΔI can become a metrological tool for an evaluation of a gate capacitance by using the following equation [11].

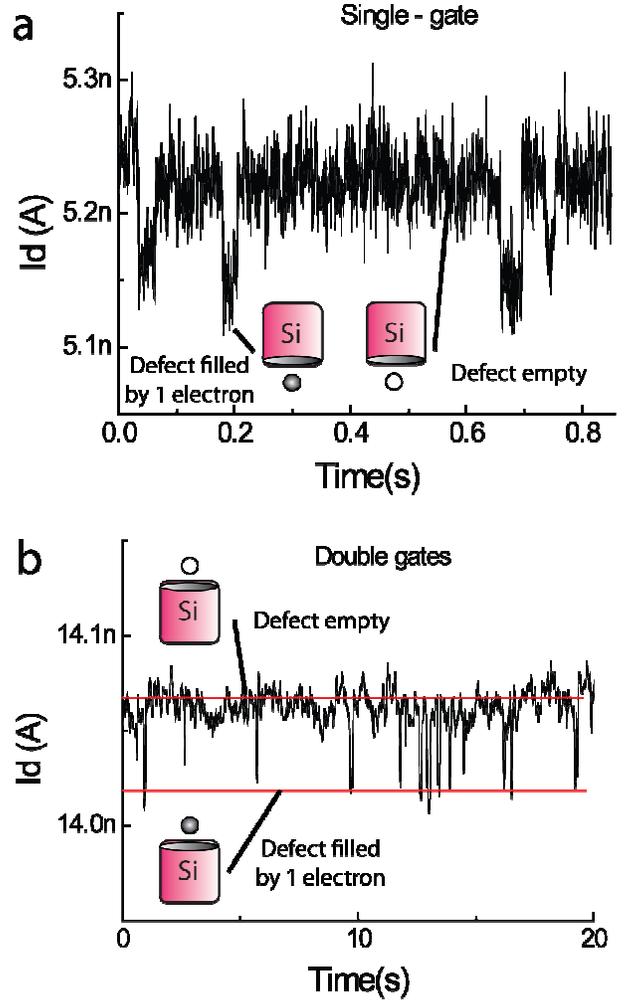

Fig. 3. 3 Estimation of capacitances from Random Telegraph Signal
(a) Examples of Random Telegraph Signal (RTS) for the single gate (back-gate) and (b) for the double gates structure.

$$\frac{\Delta I}{gm} = \frac{q*}{Cg} \qquad (2)$$

q* is an effective charge of a trapped electron. Cg stands for Cug or Cbg depending on the use of double-gate or single-gate configuration, respectively, to avoid writing twice equations. This simplicity of the expression is adapted to a transconductance gm instead of $gm_{ug} = \partial I/\partial Vug$ and $gm_{bg} = \partial I/\partial Vbg$. The q*/Cg corresponds to the shift of I-Vug or I-Vbg characteristics induced by an electron trapping. Previously, we derived that q*/q was given by $2\varepsilon_{SiO2}/(\varepsilon_{SiO2}+\varepsilon_{Si}) \approx 0.5$ in the subthreshold region and that it decreased with gate voltage due to increased screening effect caused by electrons in the channel[11] (see comments in appendix). Experimentally, an average q*/q can be obtained to 0.37 from the RTS, especially, at gate voltage corresponding to the highest transconductance. From Figs. 3a and 3b, we can evaluate $gm_{bg}$, $gm_{ug}$, $\Delta Ibg$, $\Delta Iug$ to be 1 nS, 6 nS, 90 pA, and 50 pA, respectively. As a result, we can evaluate Cbg and Cug to be 0.67 and 7.2 aF, respectively. The advantage of this technique is that $\Delta I$ and gm can be given precisely and that we do not need any assumption on device dimensions.

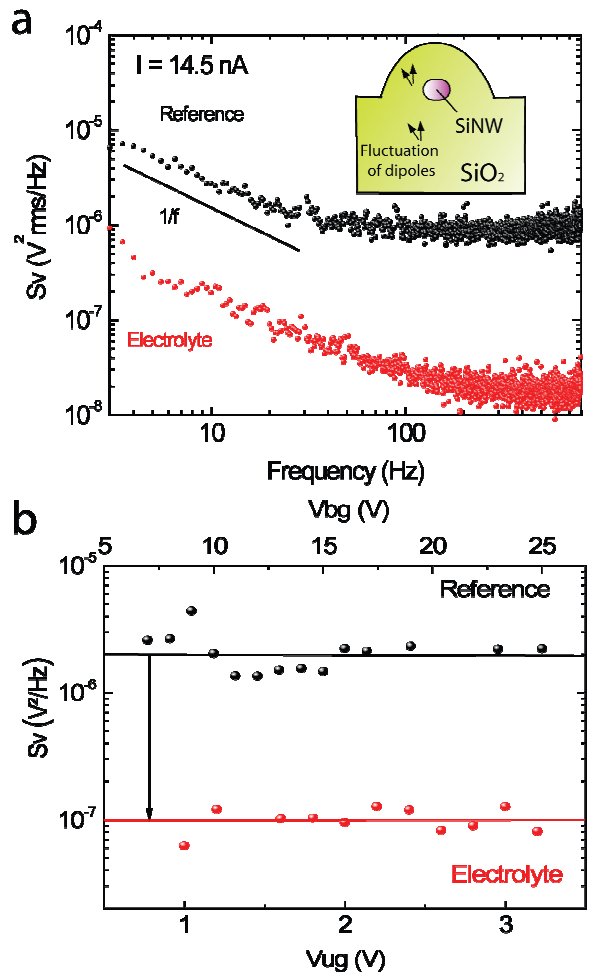

Fig. 4. Estimation of capacitances from dielectric polarization noise
(a) Sv=SI/gm² curves for the reference and liquid-gated SiNW.
(b) Vbg and Vug dependence of Sv for the reference and liquid-gated SiNW, respectively (at 10 Hz). The average amplitude is used to extract the capacitance from eq. 3.

### D. Dielectric Polarization noise

The third approach to evaluate the capacitance uses an analysis of the dielectric polarization noise[13,18]. Eq. 3 is obtained from fluctuation-dissipation theorem, which extends the Johnson-Nyquist noise equation to complex impedance[19] (see supplementary information, methods). It concerns only ultra-low noise devices (as in this work) and is due to the thermal fluctuation of dipoles in the oxide as illustrated in Fig. 4a inset (cross-section view of the device).

$$Cg = \frac{2kT\, tg\delta}{\pi Sv\, f} \quad (3)$$

Sv is a power spectrum density of voltage noise referred to the gate. It is obtained experimentally from $S_I/gm^2$ where $S_I$ is a measured power spectrum density of current noise. k is the Boltzman constant, T the temperature, and tg δ the dielectric loss of the gate oxide. The typical value[20] of tg δ of $SiO_2$ is $3.8\times10^{-3}$. The Sv of the SiNW in the single- and double-gate configurations are shown in Fig. 4a. They follow 1/f law and flat characteristics originating from Johnson-Nyquist noise. As expected from Eq. 3, Sv in the 1/f range at lower frequency is independent of gate voltage (Fig. 4b). From Eq. 3 using the average values of Sv indicated by lines in Fig. 4b over Vug and Vbg, Cug and Cbg are evaluated to be 9.68 and 0.48 aF, respectively, .

### E. Single-electron-transistor at low temperature

The fourth experimental approach for a capacitance evaluation is based on SET characteristics at low temperature. Due to Coulomb blockade originating from a small structure of the Si NW, an oscillation in I-Vg (Vug or

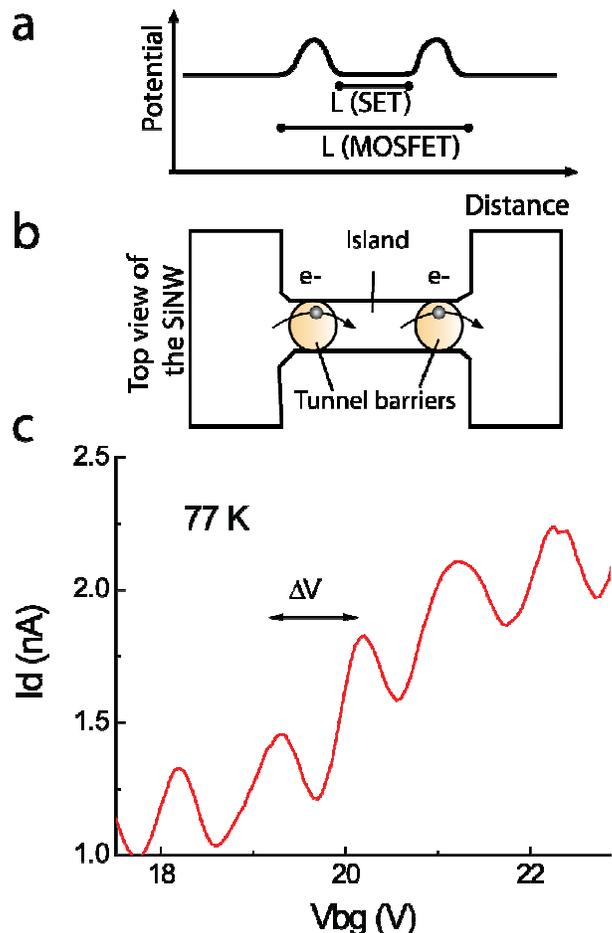

Fig. 5. Estimation of capacitances from single-electron transistor characteristics (a) Schematic view of electric potential along the contact pads and SiNW. At room temperature, electrons pass over the potential barriers which is not the case at low temperature and therefore the SiNW length L(SET) is expected to decrease at low temperature. (b) Schematic top view of the SiNW acting as a SET. Tunnel barriers arise at Source/drain and electrons transit 1 by 1 through the island at the center of the SiNW. (c) Example of SET characteristic at 77 K for the single-gate device. Capacitance is extracted from eq.4.

Vbg) characteristics with a period of ΔVg given by Eq. 4 was observed.

$$Cg = \frac{q}{\Delta Vg} \quad (4)$$

From Fig. 5c and Eq. 4, we can evaluate Cbg to be 0.2 aF with a metrology precision. Although Cug composed of the electrolytic gate cannot be evaluated at low temperature, we can estimate it to be 2 aF from the devices with the same structure except for a polysilicon upper gate[21], which is also consistent with the ratio Cug/Cbg obtained previously.

## II. DISCUSSION

### A. Summary of all techniques

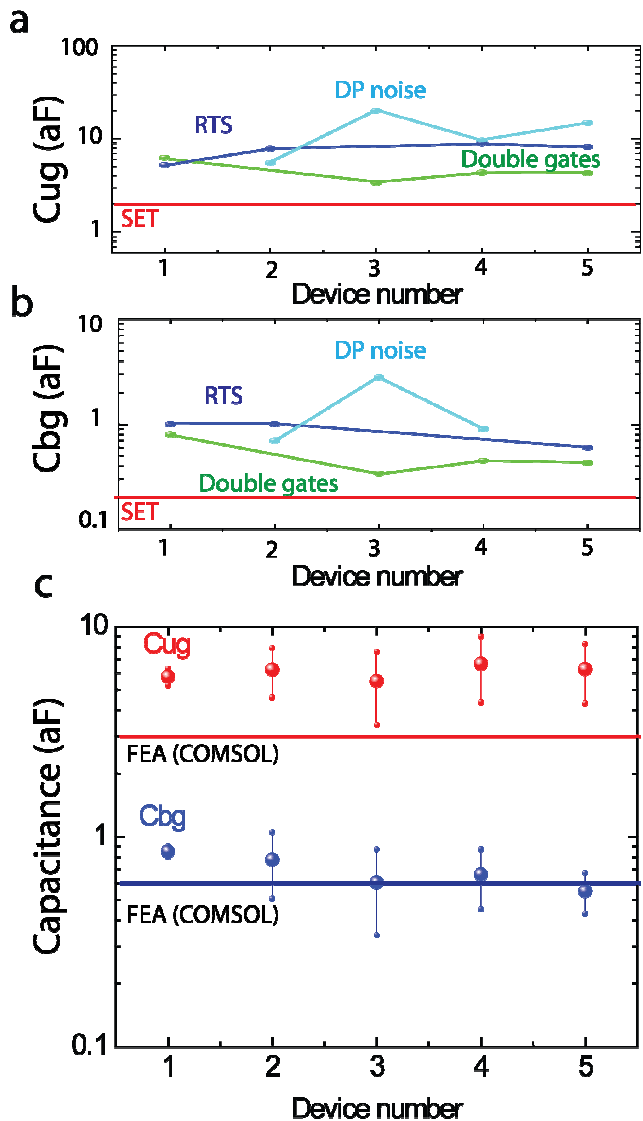

We have used these four approaches to evaluate gate capacitance of five CHEMFETs as summarized in Figs. 6a and 6b. All evaluated values for all devices are within an order of magnitude fairly acceptable, which supports that all approaches are feasible and useful. The comparison with the FEA methods supports also that feasibility as shown in Fig. 6c.

### B. Estimation of the mobility

Next, we evaluated carrier mobility in a 50-nm-long Si NW by using gate capacitance obtained above, considering the average between RTS and double gates technique. Fig. 6d shows I-Vbg and I-Vug characteristics without and with electrolyte in the micro-bath, respectively. The electrolyte was NaCl droplet whose concentration was varied from 100 μM to 1M. Fig. 6e shows enlarged I-Vug characteristics for clarity of the effect of NaCl concentration. The difference in slopes between with and without electrolyte is related to the difference in capacitances between back-gate and front-gate oxide as explained before. Good fits of the I-V curves were obtained using a classical equation (Eq. 5) of MOSFETs in the linear regime (Vd = 50 mV) and averaged gate capacitance as shown in Fig. 6c

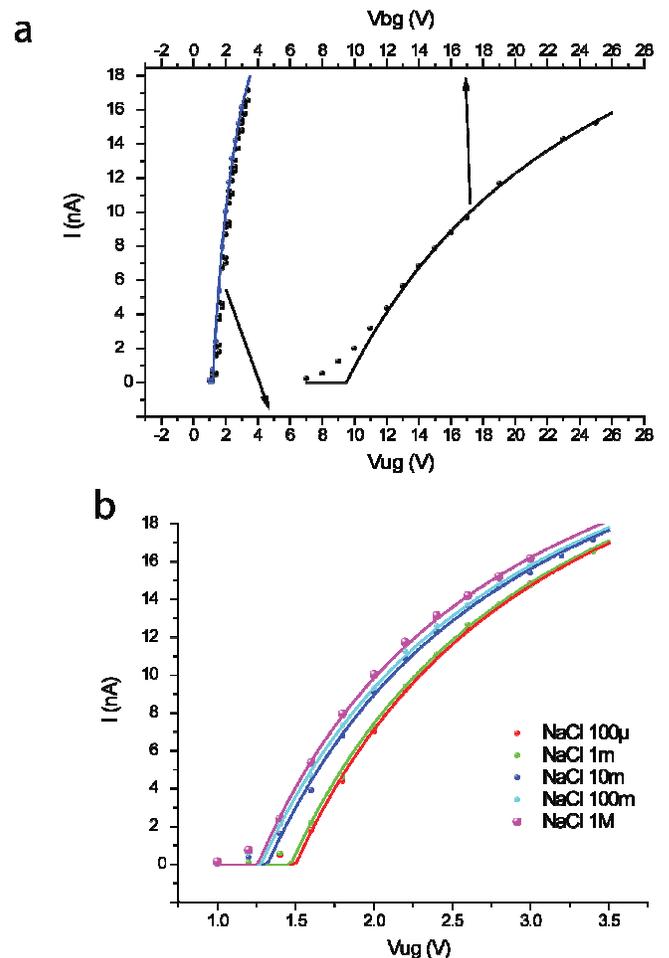

Fig. 6. Comparison of estimated capacitances from all techniques and estimation of electrons mobility (a) Cug and (b) Cbg evaluated from the different techniques. For the SET-based method, only a typical Cbg and Cug are plotted because clear oscillations are not observed for all devices in single-gate configuration and also because electrolytic gate can't be used at low temperature. Similar samples with polysilicon gates were used to estimate Cug[20]. (c) Average capacitances obtained considering corrected RTS equation as an upper limit and double-gates I-V technique as a lower limit. Finite element analysis simulation is added as a comparison.

Fig. 7. (a) I-Vug and I-Vbg in single-gate and double-gates configurations. (b) Zoom of (a). Same capacitances and mobilities are used to fit curves for different ionic strengths in liquid-gated SiNWs.

$$I = \frac{\mu C_g}{L^2}(V_g - V_{th})V_d \quad \text{with} \quad \mu = \frac{\mu_0}{1 + \theta^{-1}(V_g - V_{th})} \quad (5)$$

μ is the mobility, $\mu_0$ the mobility at low transverse electric field and Θ is a constant to account for the reduction of mobility with transverse electric field. Good fits gives μ of 74 cm²/V.s and Θ of 17 V in single-gate configuration (without electrolyte), and μ of 71.5 cm²/V.s and Θ of 1.7 V in two gates configuration (with electrolyte). It should be noted that the mobility does not vary with NaCl concentration regardless of a shift of current characteristics according to NaCl concentration[13] and that the values of $\mu_0$ of the five devices are much smaller than that of a bulk Si as shown in Table 1. Since the fluctuation of mobility among the five devices has direct correlation to neither DP nor RT noise, they would not be at the origin of the reduction in the mobility. Possible reasons why the mobility in a Si NW is so small are that there are potential fluctuation in the Si NW or that electrons scatter at the surface of the Si NW.

## III. CONCLUSION

In summary, we have used several techniques to evaluate sub-atto-farad gate capacitance of a CHEMFET with a Si NW channel. A fluctuation of gate capacitance given by each technique is within 33 %, which supports that all techniques are feasible. From evaluated gate capacitance of the CHEMFET with a 15-nm-width and 50-nm-length channel, electron mobility in the channel at low transverse field is estimated to be 68 cm²/V.s. This study gives a reference in mobility for other studies and opens door to a metrological study of Si-NW CHEMFETs used for electrical detection of a small number of molecules.

## APPENDIX

*Device fabrication*

The nanoscale MOSFETs were fabricated on a silicon-on-insulator (SOI) wafer. First, a narrow constriction sandwiched between two wider (400-nm-wide) channels was patterned on the 30-nm-thick top silicon layer (p-type, boron concentration of 10¹⁵ cm-3). The length and width of the constriction channel was 30 and 60 nm, respectively (Fig.1c). The patterning was followed by thermal oxidation at 1000°C to form a 40-nm-thick SiO2 layer around the channel. This oxidation process reduced the size of the constriction to about 15 nm, giving a final channel dimension of 15 x 50 nm. Then, we implanted phosphorous ions outside the constriction, five micrometer away from it using a resist mask, to form highly doped source and drain regions. Aluminum electrodes were evaporated on these source and drain regions. Finally, 100 nm thick Pt electrodes were evaporated using a resist mask and 300 µm thick micro-baths were also fabricated by optical lithography before a polymerization stage at 180°C for 2 hours.

*Electrical measurements*

Electrical measurements were performed at room temperature in a glove-box with a controlled N2 atmosphere (< 1 ppm of $O_2$ and $H_2O$). For noise measurements, drain voltage $V_D$ (usually 50 mV) and back-gate voltage (< 8V) were applied with an ultralow-noise DC power supply (Shibasoku PA15A1 when $V_{BG}$ < 8 V or Yokogawa 7651 when $V_{BG}$ > 8 V). The source current was amplified with a DL 1211 current preamplifier supplied with batteries. RTS data and noise spectra were acquired with an Agilent 35670 dynamic signal analyzer. A detailed protocol is published in ref [22]. For standard I-V characteristics, a semiconductor parameter analyzer agilent 4156C was used.

*Trap effective charge q* in Eq.2 (RTS)*

When a charge q, located in a material 1 (e.g.SiO2) with a relative permittivitty ε1, close to the interface of a material 2 (e.g.Si) with a different relative permittivity ε2, an image charge q(ε1-ε2)/(ε1+ε2) exists in the material 2. Therefore, far from the interface (e.g. gate), an effective trap charge q*=2qε1/(ε1+ε2), is observed. When an accumulation of electrons exists in the channel due to capacitive coupling with the gate, the dielectric constant of material 2 is increased and we get Eq. S1.

$$\frac{q^*}{q} = \frac{2\varepsilon_1}{(\varepsilon_1 + \varepsilon_2)} \left[ \frac{1}{1 + \frac{qt_{acc}}{2kT\varepsilon_0(\varepsilon_1 + \varepsilon_2)} \frac{C_G}{WL}(V_g - V_{th})} \right] \quad (6)$$

$C_G$ is the gate capacitance, tacc the accumulation layer thickness, $\varepsilon_0$ the vacuum dielectric permittivity, W and L nanowire's width and length, Vg the back-gate voltage and Vth the flat-band voltage. In a previous work[11], it was considered that the Debye screening length of the undoped silicon is very large (>> 100 nm) compared to Si thickness (15 nm thick Silicon on Insulator and therefore ε2≈ε1 in Eq. 6. This is a first order approximation valid in 2D and enough for the aim of that paper. In a 3D approach, screening can also be done by the wide silicon contact leads (400 nm) and therefore Eq. S1 should be used as is. In the case of SiO2 oxide and SiNW the prefactor in Eq. 2 is 0.5. By comparison with other technique, the 3D approach seems appropriate and therefore the dielectric constant should be taken into account for RTS amplitude analysis. An example of gate voltage dependence of q* with Vg is shown in [11].

*Dielectric polarization noise (Eq.3)*

Considering the impedance $Z=(j.2\pi.C.f)^{-1}$ of the gate capacitor with $C=Cg-jCg''$ and $tg\delta=C''/Cg$, the extension of thermal Nyquist noise for a complex impedance (fluctuation-dissipation theorem) leads to:

$$Sv = 4kT\Re(Z) = \frac{2kTC''}{\pi Cg^2} = \frac{2kT tg\delta}{\pi Cgf} \qquad (7)$$

ACKNOWLEDGMENT

We sincerely thank D. Theron for carefull reading of the manuscript and S. Lamant and J. Oden for assistance with Finite Element Analysis (COMSOL) simulation.

REFERENCES

[1] Y. Cui, Z. Zhong, D. Wang, W. U. Wang and C.M. Lieber, "High performance silicon nanowire transistors" Nano Lett. 3, 149-152 (2003)
[2] C. Thelander, P. Agarwal, S. Brongersma, J. Eymery, L. Feiner, A. Forchel, M. Scheffler, W. Riess, B.J. Ohlsson, V. Gosele, L. Samuelson "nanowire-based one-dimensional electronics" Mater.Today 9, 28-35 (2006)
[3] L. Zhang, R. Tu, H.J. Dai "Parallel Core–Shell Metal-Dielectric-Semiconductor Germanium Nanowires for High-Current Surround-Gate Field-Effect Transistors " Nanolett. 6, 2785-2789 (2006)
[4] K. Nishiguchi, H. Inokawa, Y. Ono, A. Fujiwara and Y. Takahashi, "Multilevel memory using an electrically formed single-electron box" Appl.Phys.Lett. 85, 1277-1280 (2004)
[5] F. Patolsky, G. Zheng and C.M. Lieber," Fabrication of silicon nanowire devices for ultrasensitive, label-free, real-time detection of biological and chemical species" Nat. Protocols 1, 1711-1724 (2006)
[6] K. Nishiguchi, N. Clement, Y. Yamaguchi and A. Fujiwara, Appl.Phys.Lett "Si nanowire ion-sensitive field-effect transistors with a shared floating gate". 94, 163106-160109 (2009)
[7] S. Clavaguera, A. Carella, L. Cailler, C. Celle, J. Pecaut, S. Lenfant, D. Vuillaume and J.-P. Simonato Angew.Chem.Int.Ed. "Sub-ppm Detection of Nerve Agents Using Chemically Functionalized Silicon Nanoribbon Field-Effect Transistors "49, 4053 (2010)
[8] R. Tu, L. Zhang, Y. Nishi and H. Dai, "Measuring the Capacitance of Individual Semiconductor Nanowires for Carrier Mobility Assessment " Nanolett. 7, 1561 (2007)
[9] O. Gunawan, L. Sekaric, A. Majumdar, M. Rooks, J. Appenzeller, J.W. Sleight, S. Guha and W. Haensch, "Measurement of carrier mobility in silicon nanowires" Nanolett. 8, 1566 (2008)
[10] A. Tselov, S.M. Anlage, Z. Ma and J. Meingailis "Broadband dielectric microwave microscopy on micron length scales" Rev.Sci.Inst. 78, 044701-7 (2007)
[11] N. Clement, K. Nishiguchi, A. Fujiwara and D. Vuillaume, "One by one trap activation in a silicon nanowire transistor" Nature communications doi: 10.1038/ncomms1092 (2010)
[12] A. Gokirmak, H. Inaltehin and S. Tiwari "Attofarad resolution capacitance–voltage measurement of nanometer scale field effect transistors utilizing ambient noise" Nanotechnology 20, 335203-335208 (2009)
[13] N. Clement, K. Nishiguchi, J.-F. Dufreche, D. Guerin, A. Fujiwara and D. Vuillaume, Appl.Phys.Lett. 98, 014104 (2011)
[14] A. Fujiwara, S. Horiguchi, M. Nagase and Y. Takahashi, "Threshold Voltage of Si Single-Electron Transistor" Jpn.J.Appl.Phys. 42, 2429 (2003)
[15] K. Nishiguchi, Y. Ono, A. Fujiwara, H. Inokawa and Y. Takahashi, "Stochastic data processing circuit based on single electrons using nanoscale field-effect transistors" Appl.Phys.Lett. 92, 062105 (2008); Nature 453, 166 (2008)
[16] K. Nishiguchi, C. Koechlin, Y. Ono, A. Fujiwara, H. Inokawa and H. Yamaguchi, "Single-Electron-Resolution Electrometer Based on Field-Effect Transistor" Jpn.J.Appl.Phys. 47, 8305 (2008)
[17] S. Horiguchi, A. Fujiwara, H. Inokawa, and Y. Takahashi, Jpn. J. Appl. Phys. "Analysis of Back-Gate Voltage Dependence of Threshold Voltage of Thin Silicon-on-Insulator Metal-Oxide-Semiconductor Field-Effect Transistor and Its Application to Si Single-Electron Transistor " 43, 2036 (2004).
[18] N.E. Israeloff, "Dielectric polarization noise through the glass transition" Phys.Rev.B. 53, R11913 (1996)
[19] H. Bouchiat and M. Ocio, Comments Condens. Matter Phys. 14, 163 (1988)
[20] G.G. Raju "Dielectrics in electric field", Marcel Dekker New York (2003)
[21] M. Nagase, S. Horiguchi, K. Shiraishi, A. Fujiwara, and Y. Takahashi:, "Single-electron devices formed by thermal oxidation " Journal of Electroanalytical Chemistry 559, 19-23 (2003)
[22] N. Clement, K. Nishiguchi, A. Fujiwara and D. Vuillaume "Measuring and analyzing low frequency noise in nanodevices" Protocol Exchange (2010) doi:10.1038/protex.2010.20

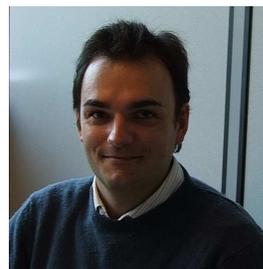

**Nicolas Clement** was born in 1977. He received the Electronics Engineer degree from the Institut Supérieur d'Electronique et du Numérique, Toulon, France, 2000 and the PhD degree in solid-state physics, from the University of Marseille, France in 2003. From 2003 to 2005, he was a post doc researcher at the Nippon Telegraph and Telephone Corporation (NTT), Japan. He is actually a researcher at CNRS (centre national de la recherche scientifique) and he works at the Institute for Electronics, Microelectronics and Nanotechnology (IEMN), University of Lille. He is in the « Molecular Nanostructures & Devices » research group at IEMN held by Dominique Vuillaume. His research interests are nanoscale devices, metrology, noise, molecular electronics and chemical field effect transistors.

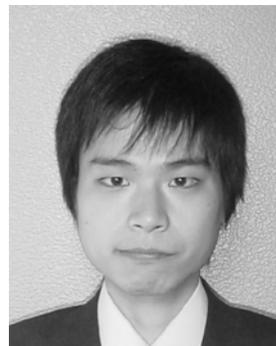

**Katsuhiko Nishiguchi** received the B.E., M.E., and Ph. D in electrical engineering in 1998, 2000, and 2002, respectively, from Tokyo Institute of Technology, Tokyo, Japan. Since joining Nippon Telegraph and Telephone Corporation (NTT) in 2002, he has been engaged in the research on physics and technology of Si nanometer scale devices for LSI applications. He is a senior research scientist. He received IUPAP Young Author Best Paper Award at the ICPS meeting, Graduate Student Award Silver at the MRS meeting, and Young Scientist Award at the JSAP meeting in 2000. Dr. Nishiguchi is a member of the Japan Society of Applied Physics.

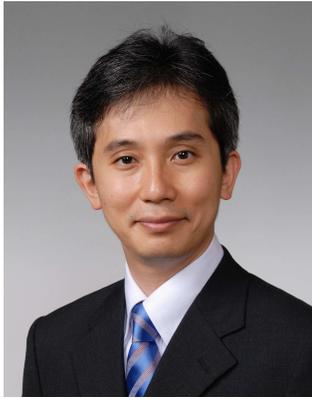

**Akira Fujiwara** was born in Japan on March 9, 1967. He received the B.S., M.S., and Ph.D. degrees in applied physics from The University of Tokyo, Japan, in 1989, 1991, and 1994, respectively. In 1994, he joined LSI Laboratories, Nippon Telegraph and Telephone (NTT) Corporation, Kanagawa, Japan. He moved to the Basic Research Laboratories (BRL) in 1996. Since 1994, he has been engaged in research on silicon nanostructures and their application to single-electron devices. He was a guest researcher at National Institute of Standards and Technology (NIST), Gaithersburg, USA during 2003-2004. Since 2006, He is a group leader of Nanodevcies Research Group, NTT BRL. Since 2007, He is a Distinguished Technical Member, NTT BRL. He received SSDM Young Researcher Award in 1998, SSDM Paper Award in 1999, and Japanese Journal of Applied Physics (JJAP) Paper Awards in 2003 and 2006. He was awarded Young Scientist Award from the Minister of MEXT (Ministry of Education, culture, sports, science, and technology) in 2006. He is a member of the Japan Society of Applied Physics and IEEE.

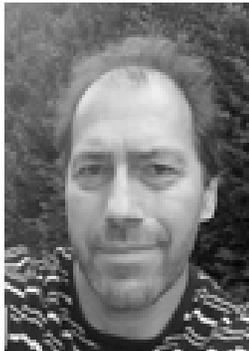

**Dominique Vuillaume** was born in 1956. He received the Electronics Engineer degree from the Institut Supérieur d'Electronique du Nord, Lille, France, 1981 and the PhD degree and Habilitation diploma in solid-state physics, from the University of Lille, France in 1984 and 1992, respectively. He is research director at CNRS (centre national de la recherche scientifique) and he works at the Institute for Electronics, Microelectronics and Nanotechnology (IEMN), University of Lille. He is head of the « Molecular Nanostructures & Devices » research group at IEMN.

His research interests (1982-1992) covered physics and characterization of point defects in semiconductors and MIS devices, physics and reliability of thin insulating films, hot-carrier effects in MOSFET's. Since 1992, he has been engaged in the field of Molecular Electronics. His current research concerns:

- design and characterization of molecular and nanoscale electronic devices,
- elucidation of fundamental electronic properties of these molecular and nanoscale devices,
- study of functional molecular devices and integrated molecular systems,
- exploration of new computing paradigms using molecules and nanostructures.

He is the author or co-author of 150 scientific (peer-reviewed) papers in these fields. He was scientific advisor for industrial companies (Bull R&D center) and he is currently scientific advisor for the CEA "Chimtronique" research program.